\documentclass{sig-alternate}

\usepackage{amsmath}
\usepackage{lipsum}
\usepackage{graphicx, subfigure}
\usepackage{booktabs}
\usepackage[hyphens]{url}

\usepackage{hyperref}

\newtheorem{definition}{Definition}

\newcommand{\Q}{\mathbf{Q}}
\newcommand{\D}{\mathbf{D}}
\newcommand{\C}{\mathbf{C}}
\newcommand{\Lb}{\mathbf{L}}

\newcommand{\U}{\mathbf{U}}

\makeatletter
\def\@copyrightspace{\relax}
\makeatother

\begin{document}


\sloppy

\title{Context Models For Web Search Personalization}

\numberofauthors{3}

\author{
\alignauthor Maksims N.~Volkovs\\
       \affaddr{University of Toronto}\\
       \affaddr{40 St.~George Street}\\
       \affaddr{Toronto, ON M5S 2E4}\\
       \email{mvolkovs@cs.toronto.edu}
}

\maketitle

\begin{abstract}
We present our solution to the Yandex Personalized Web Search 
Challenge. The aim of this challenge was to use the historical 
search logs to personalize top-N document rankings for a set 
of test users. We used over 100 features extracted 
from user- and query-depended contexts to train neural net
and tree-based learning-to-rank and regression models. Our
final submission, which was a blend of several different models, 
achieved an NDCG@10 of 0.80476 and placed 4'th amongst the
194 teams winning 3'rd prize\footnote{Top team ``pampampampam" was from
Yandex and did not officially participate in the competition.}.
\end{abstract}

\section{Introduction}

Personalized web search has recently been receiving a lot of 
attention from the IR community. The traditional 
one-ranking-for-all approach to search often fails for
ambiguous queries (e.g. ``jaguar") that can refer to 
multiple entities. For such queries, non-personalized 
search engines typically try to retrieve a diverse set 
of results covering as many possible query
interpretations as possible. This can result in highly 
suboptimal search sessions, where web pages that the user is 
looking for are very low in the returned ranking.

In many such cases previous user search history
can help resolve the ambiguity and personalize (re-rank)
returned results to user-specific information needs. Recently, 
a number of approaches have shown that 
search logs can be effectively mined to learn accurate personalization
models \cite{SATClick1, SATClick2, SATClick3, Fatures1, Fatures2},
which can then be deployed to personalize retrieved results 
in real time. Many of these models do not require any external 
information, and obtain all learning signals directly 
from the search logs. Such models are particularly effective 
since search logs can be collected at virtually no cost to the 
search engine, and most search engines already collect them by default.
\begin{figure}[t]
\centerline{
	\includegraphics[scale=0.31]{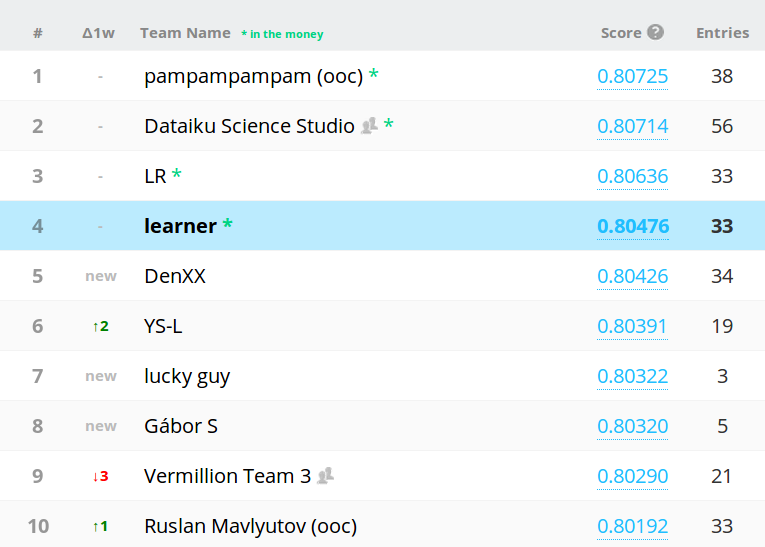}
}
\caption{Final leaderboard standings, our team ``learner" placed 4'th amongst 
the 194 teams (261 users) that participated in this challenge.}
\label{fig:leaderboard}
\end{figure}

To encourage further research in this area Yandex recently partnered 
with Kaggle and organized the Personalized Web Search Challenge\footnote{\url{www.kaggle.com/c/yandex-personalized-web-search-challenge}}. 
At the core of this challenge was a large scale search log dataset
released by Yandex containing over 160M search records. The goal
of the challenge was to use these logs to personalize
search results for a selected subset of test users. In this 
report we describe our approach to this problem.
The rest of the paper is organized as follows, Section 
\ref{sc:challenge} describes the challenge data and task in detail.
Section \ref{sc:our_approach} introduces our approach in three stages:
(1) data partitioning, (2) feature extraction and (3) model training. 
Section \ref{sc:results} concludes with results.

\section{Challenge Description}\label{sc:challenge}

In this challenge Yandex provided a month's (30 days) worth of search engine logs for a 
set of users $\U = \{u_1,...,u_N\}$. Each user $u$ engaged with the search engine
by issuing queries $\Q_u = \{q_{u1},...,q_{uM_u}\}$. Queries that were issued 
``close" to each other in time were grouped into sessions. For each query 
$q_u$ the search engine retrieved a ranked list of web pages 
(documents) $\D_{q_u} = \{d_{q_u1},...,d_{q_uK_{q_u}}\}$, returning it to the user.
User then scanned this list (possibly) clicking on some documents. Every such click
is recorded in the logs together with time stamp and id of the document that
was clicked. Only the top ten documents and their clicks (if any) 
were released for each query so $K_{q_u} = 10$ $\forall q_u$. For privacy 
reasons, very little information about queries and documents was provided. 
For queries, only numeric query id and numeric query-term ids were released. 
Similarly, for documents, only numeric document id and corresponding domain id 
(i.e. facebook.com for facebook pages) were released.

Clicks combined with dwell time (time spent on a page) can provide a good 
indication of document relevance to the user. In particular, it has been consistently
found that longer dwell times strongly correlate with high relevance
leading to the concept of {\it satisfied} (SAT) clicks -- clicks with dwell time 
longer than a predefined threshold (for example 30 seconds)~\cite{SATClick1, SATClick2, SATClick3}. Most existing personalization frameworks assume that documents with
SAT clicks are relevant and use them to train/evaluate models.

This competition adopted a similar evaluation framework where each document was assigned 
one of three relevance labels depending on whether it was clicked and click dwell 
time length. For privacy reasons dwell time was converted into anonymous 
``time units" and relevance labels were assigned according to the following criteria:
\begin{itemize}
\item {\bf relevance 0}: documents with no clicks or dwell time strictly less than 
50 time units

\item {\bf relevance 1}: documents with clicks and dwell time between 50 and 
399 time units

\item {\bf relevance 2}: documents with clicks and dwell time of at least 
400 time units as well as documents with last click in session
\end{itemize}
Using above criteria, a set of relevance labels 
$\Lb_{q_u} = \{l_{q_u1},...,l_{q_uK_u}\}$ (one per document) can be generated for 
every issued query. Note that these relevance labels are personalized to the user who 
issued the query and express his/her preference over the returned documents. Given 
the relevance labels, the aim of the challenge was to develop a personalization 
model which would accurately re-rank the documents in the order of relevance to 
the user who issued the query.

\begin{figure*}[t]
\centerline{
	\includegraphics[scale=0.55]{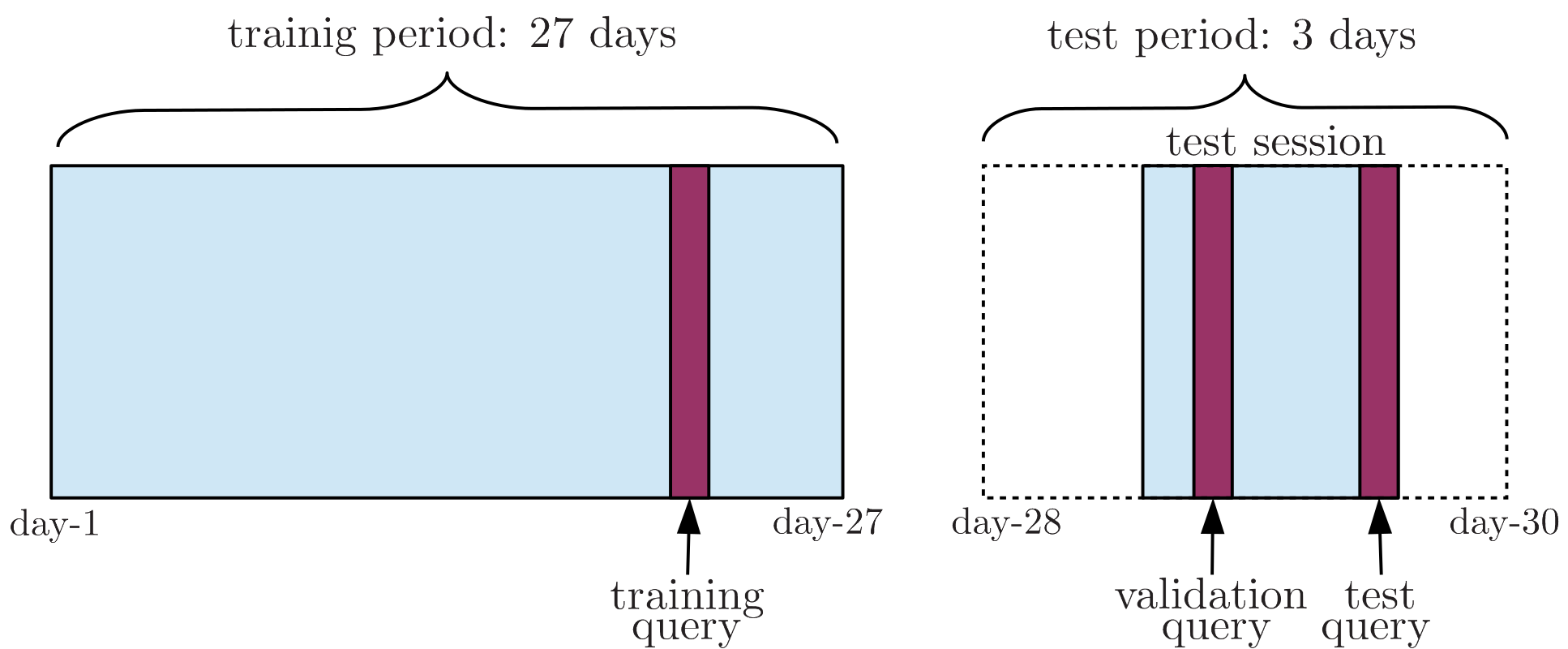}
}
\caption{Diagram showing data partition and training/validation/test query selection 
(in red) for a {\it single} user. Training query was always selected to be the
last query in training period with at least one relevant (relevance > 0) document. 
Similarly, validation query was always selected to be the last query in test session 
with at least one relevant document. Test query was given a priori and was the last 
query in test session.}
\label{fig:data_partitioning}
\end{figure*}

To ensure fair evaluation the data was partitioned into training
and test sets. Training data consisted of all queries issued in the first 27 
days of search activity. Test data consisted of queries sampled from the next
3 days of search activity. To generate the test data one query with at least one 
relevant (relevance > 0) document was sampled from 797,867 users resulting in a 
fairly large test set with almost 800K queries and 8M documents. In order to simulate 
real-time search personalization scenario, all search activity after each test 
query was removed from the data. Furthermore, to encourage discovery of medium 
and long term search preference correlations all sessions except those that contained 
test queries were removed from the 3 day test periods. A diagram illustrating 
data partitioning is shown in Figure \ref{fig:data_partitioning}, and full dataset statistics are shown in Table \ref{tb:dataset_stats}.
\begin{table}[t]\centering
\caption{Dataset statistics}
\begin{tabular}{lr}
\toprule
Unique queries 					&21,073,569\\
Unique documents 				&70,348,426\\
Unique users 					&5,736,333\\
Training sessions 				&34,573,630\\
Test sessions 					&797,867\\
Clicks in the training data 		&64,693,054\\
Total records in the log 		&167,413,039\\
\bottomrule
\end{tabular}
\label{tb:dataset_stats}
\end{table}

All submissions were required to provide full document rankings for each of the
797,867 test queries and were evaluated using the Normalized Discounter Cumulative Gain 
(NDCG) \cite{NDCG} objective. Given a test query $q_u$ with documents $\D_{q_u}$ 
and relevance labels $\Lb_{q_u}$, NDCG is defined by:
\begin{equation}
 \mbox{NDCG}(\pi, \Lb_{q_u})@T = \frac{1}{G_T(\Lb)} 
 \sum_{i=1}^T \frac{2^{\Lb(\pi^{-1}(i))} -
1}{\log_2(i + 1)}
\end{equation}
Here $\pi : \D_{q_u} \to \{1,...,M_u\}$ is a ranking produced by the model
mapping each document $d_{q_uj}$ to its rank $\pi(j) = i$, and 
$j = \pi^{-1}(i)$. $\Lb(\pi^{-1}(i))$ is the relevance label of the document 
in position $i$ in $\pi$, and $G_T(\Lb)$ is a normalizing constant. Finally,
$T$ is a truncation constant which was set to 10 in this challenge.

As commonly done in data mining challenges, test relevance labels were not 
released to the participants and all submission were internally evaluated
by Kaggle. Average NDCG@10 accuracies for approximately 50\% of test queries 
were made visible throughout the challenge on the ``public" leaderboard
while the other 50\% were used to calculate the final standings
(``private" leaderboard). 
\begin{table}[t]\centering
\caption{Document relevance distribution for training and validation sets.}
\begin{tabular}{lcc}
\toprule
\phantom{ab} & \multicolumn{1}{c}{Training} &
\multicolumn{1}{c}{Validation}\\
\midrule
no click&		5,673,937&	1,993,602\\
relevance 0&		115,713&		54,572\\
relevance 1&		206,658&		196,290\\
relevance 2&		728,662&		149,536\\
\bottomrule
\end{tabular}
\label{tb:relevance_stats}
\end{table}

\section{Our Approach}\label{sc:our_approach}

In this section we describe our approach to this challenge. 
Before developing our models we surveyed existing work in this
area and found that most personalization methods can be divided into three
categories: heuristic, feature-based and user-based. Heuristic methods
\cite{ClickBaseline} use search logs to compute user-specific 
document statistic, such as the number of historical clicks, 
and then use this statistic to re-rank the documents. Since
it is often difficult to know which statistic will work best,
feature-based models \cite{Fatures1, Fatures3, Fatures2} extract
a diverse set of features used as input for machine learning methods 
that automatically learn personalization models. Note that while
features are extracted separately for every user-query-document 
triplet, the same model is used to re-rank documents for all users.

Finally, user-based methods \cite{CubeSVD, UserTopicPageRank, FeatureRerank}
as the name suggests, learn separate models for each user. Some of 
these models use collaborative filtering techniques to infer
latent factors for users and documents \cite{CubeSVD, UserTopicPageRank}, 
while others adapt learning-to-rank models by incorporating 
user-specific weights and biases \cite{FeatureRerank}.

User-based models allow the highest level of personalization but 
require extensive user search history and/or side information about
queries and documents (such as topics, document features etc.). Given the 
sparsity of our data (70M unique documents in 160M records) and 
lack of user/query/document information we opted to use the 
feature-based approach. In the following sections we 
describe in detail all the components that were necessary to create a 
feature-based model, namely data partitioning, feature extraction and
learning/inference algorithms.

\subsection{Dataset Partitioning}

We begin by describing our data partitioning strategy. Properly selected
training/validation datasets are crucial to the success of any data mining model. 
Ideally we want these datasets to have very similar properties to the test data.
To achieve this we carefully followed the query sampling procedure (described
in Section \ref{sc:challenge}) used by competition administrators to select  
test queries.

For each user we first sorted all sessions by day (lowest to highest) randomly
resolving ties since exact timestamps were not available. We then selected 
the last query in the 27 day training period with at least one relevant (relevance > 0) 
document as training query. Similarly, last query in test session with
at least one relevant document was selected for validation. This selection 
process is shown in Figure \ref{fig:data_partitioning}. 

The motivation behind choosing these specific queries was 3-fold. First, since 
features can only be extracted from queries issued {\it before} the given query, we
need to choose queries with as much historical data as possible. Selecting 
queries at the end of training period and test session ensures maximum
historical data. Second, there could be a large time gap between the end of 
training period and test session, and during that time the user's search needs and 
preferences could change significantly. To capture this we need both  
training and validation queries to be as close as possible to test ones. 
However, since many test session did not have enough data to
select two queries, only validation query was sampled from this session.
Finally, only selecting queries with at least one relevant document ensures that 
their is sufficient training signal for learning-to-rank models. Training objectives 
in these models are often order-based and thus require at least one relevant 
document.

Applying this procedure to each of the 797,867 test users and removing 
users that did not have enough data, resulted in 672,497 training and 239,400
validation queries. Once the data was partitioned relevance labels were computed for 
all documents in both training and validation queries using the criteria outlined in 
Section \ref{sc:challenge}. Table \ref{tb:relevance_stats} shows relevance
label distribution across documents in both sets. From this table we see
that the majority of documents with clicks have relevance label 1 or 2. This 
suggests that once the user clicks on a document (s)he tends to spend ``significant" 
amount of time going through the content of that document. It can also be seen 
that validation relevance distribution is similar to training one with the 
exception that training data has considerably more highly relevant 
(relevance 2) documents.

\subsection{Feature Extraction}\label{sc:features}

After partitioning the data and computing relevance labels we proceeded to 
feature extraction. Our aim was to extract features for every training, validation 
and test user-query-document triplet $(u, q_u, d_{q_u})$.
As mentioned above, the available log data provided very little information 
about individual queries and retrieved documents. For queries, we only had 
access to term vectors with individual terms converted to numeric ids. Similarly, for 
documents we only had access to their domain ids and base ranking generated by the
Yandex search engine. In this form the personalization problem is similar to 
collaborative filtering/ranking where very little information about items 
and users is typically available. Neighborhood-based models that extract
features from similar items/users have been shown to consistently perform
well in these problems and were an essential part of the Netflix prize winning
solution \cite{Netflix}. In search personalization, ranking models learned
on features extracted from user's search neighborhoods (historical sessions, 
queries etc.) have also been recently shown to perform well 
\cite{Fatures1, Fatures2, Fatures3}. Inspired by these results we 
concentrated our efforts on designing features using historical search
information in the logs.

We began by identifying several ``contexts" of interest. Here, contexts are 
analogous to user/item neighborhoods in collaborative filtering,
and contain collections of queries that have some relation to the target user-query-document triplet for which the features are being extracted. 
Formally we define context as:
\begin{definition}{\ } \\ 
Context 
$\C = \{\{q_1,...,q_M\}, \{\D_{q_1},...,\D_{q_M}\}, \{\Lb_{q_1},...,\Lb_{q_M}\}\}$
is a set of queries with corresponding document and relevance label lists.
\end{definition}

Given a user-query-document triplet $(u, q_u, d_{q_u})$, we primarily 
investigated two context types: user-related and query-related. For
user-related contexts we considered all queries issued by $u$ before 
$q_u$ and partitioned them into 2 contexts - repetitions of $q_u$ and 
everything else. The rationale behind this partitioning is that
past instances of $q_u$ are particularly useful for inferring
user's search interests for $q_u$ \cite{ClickBaseline}, and should 
be processed separately. In addition to 
historical queries from $u$, we computed context from all instances of
$q_u$ issued by users other than $u$. This context provides global
information on user preferences for documents in $q_u$, and can be
useful when little information from $u$ is available.

For each of these contexts we computed features on both document and 
domain levels. To use domains we simply substituted $d_{q_u}$ with its domain
and replaced document lists in each context with domain lists.
Given that multiple documents can have the same domain we expect 
domain features to be less precise. However, domain data
is considerably less sparse ($\sim$70M unique documents vs $\sim$5M unique domains) 
and can thus provide greater coverage. Using both document and domain lists 
we ended with a total of 6 contexts:
\begin{itemize}
\item $\C_1$: all repetitions of $q_u$ by $u$
\item $\C_2$: same as $\C_1$ but with domain lists
\item $\C_3$: all queries other than $q_u$ issued by $u$
\item $\C_4$: same as $\C_3$ but with domain lists
\item $\C_5$: all repetitions of $q_u$ by users other than $u$
\item $\C_6$: same as $\C_5$ but with domain lists
\end{itemize}
In this form our contexts are similar to ``views" explored in 
\cite{Fatures1}. The main difference between the two is that views are 
user-specific whereas contexts can include any set of queries including
those from other users. Note that we also do not apply any session-based partitioning 
within the contexts and all queries are simply aggregated together. 
Throughout the challenge we experimented with several session-related contexts 
(current session vs past sessions) but did not find them to give significant 
improvement.

After specifying the contexts we defined a total of 20 context-dependent features 
described in detail in Appendix \ref{app:context_features}. Most of these 
features aim to capture how frequently $d_{q_u}$ was shown/clicked/skipped/missed in
the given context. The features also try to account for the rank position of 
$d_{q_u}$ across the context and similarity between $q_u$ and context
queries. Query similarity features $g_4$ - $g_9$ (see Appendix \ref{app:context_features}) are only relevant when queries other than $q_u$ are included in the context, and
are thus only extracted for contexts $\C_3$ and $\C_4$. All together, we computed
$20$ features for $\C_1$, $\C_2$, $\C_5$, $\C_6$ and $16$ features for $\C_3$, 
$\C_4$ giving us a total of $112$ context features. In addition to these
features, we added rank of $d_{q_u}$ returned by the search engine as the 
$113$'th and final feature.

All of the $20$ context features only require simple operations and are straightforward 
to implement. Similarly, contexts $\C_1$ - $\C_4$ are readily available in the log data
and can be easily extracted. Contexts $\C_5$ and $\C_6$ on the other hand, are trickier
to compute efficiently since they require access to all instances of a particular
query. To calculate these we created an inverted hash map index mapping each unique 
query id to a table storing all occurrences of this query id in the logs with
corresponding document, domain and relevance label lists. For any query a single
lookup in this index was then required to compute features for every document 
returned for that query. The full features extraction for training, validation and test
queries ($\sim$1.7M queries with 17M documents) implemented in Matlab took 
roughly 7 hours on a Thinkpad W530 laptop with Intel i7-3720QM 2.6 GHz processor and 
32GB of RAM.

\subsection{Learning and Inference}

We trained several learning-to-rank and regression models on the extracted feature
data. For learning-to-rank models we used RankNet \cite{RankNet}, 
ListNet \cite{ListNet} and a variation of BoltzRank \cite{BoltzRank}. Given the 
success of tree-based generalized gradient boosting machines (GBMs) on recent IR 
benchmarks such as the Yahoo!'s Learning To Rank challenge \cite{Yahoo_LTR}, we 
also experimented with state-of-the-art GBM learning-to-rank model 
LambdaMART \cite{LambdaMART}. We omit the details of each model in this report 
and refer the reader to respective papers for detailed descriptions.

For pairwise RankNet model we experimented with various ways to extract 
pairwise preferences from click data. Specifically, many studies have
shown that users scan returned results from top to bottom \cite{topDownUser, ERR}
so documents ranked below the bottom-most click were likely missed
by the user. It is thus unclear whether we should use those documents during
training and if so what relevance should they be assigned. Skipped documents 
(i.e. those above the bottom-most click) on the other hand, were clearly 
found not relevant by the user. However, it is also unclear whether they 
should be assigned the same relevance label 0 that is given to clicked documents 
with low dwell time. Intuitively, it seems like click is a stronger preference 
signal than skip even if dwell time after that click is low.

To validate these hypotheses, we used a 1-hidden layer neural net 
implementation of RankNet and trained it on different preference targets 
extracted from clicks. We experimented with several variations of the 
cascade click model \cite{topDownUser} as well as various relevance 
re-weightings. Across these experiments the best results were obtained 
by simply setting relevance of skipped and missed documents to zero and training
on all the available data. These results, although somewhat surprising, 
can be possibly explained by the fact that this assignment matches the 
target one used in NDCG for model evaluation. In light of these results we 
used the $\{0,1,2\}$ relevance assignment in all subsequent experiments.

\section{Results}\label{sc:results}

In this section we describe the main results achieved by our models. 
Throughout the experiments we consistently found that performance 
(gains/losses) on our in house validation set closely matched the public 
leaderboard. At the end of the competition we also saw that public
and private leaderboard results were very consistent. In this 
report we thus concentrate on private leaderboard NDCG scores since
these scores were used to compute the final standings.
We note that these results were only available {\it after} the
competition ended so it was impossible to directly optimize the
models for this set.

At the beginning of the competition, before applying sophisticated
machine learning methods, we created a simple heuristic-based model 
that re-ranked documents based on their total historical relevance. 
Specifically, for every test document $d_{q_u}$ we computed
feature $g_1$ (see Appendix \ref{app:context_features}) 
using all previous instances of $q_u$ issued by $u$ (context $\C_1$).
We then re-ranked documents by $g_1$\footnote{We also experimented with features 
$g_2$ - $g_4$ but found $g_1$ to work best.} 
using original ranking to resolve ties. This model produced an NDCG@10 of 
$0.79754$ shown in Table \ref{tb:results} (``re-rank by hist relevance") which 
is a relative improvement of 0.0062 over the baseline non-personalized 
ranking produced by Yandex's search engine. This submission would have placed 
$32$'nd on the final leaderboard.
\begin{table}[t]\centering
\caption{Private leaderboard average NDCG@10 results. Only
results for the best model of each type are shown.}
\begin{tabular}{lc}
\toprule
\multicolumn{1}{c}{Model} &
\multicolumn{1}{c}{NDCG@10}\\
\midrule
default ranking	baseline			&0.79133\\
re-rank by hist relevance		&0.79754\\
regression (NN)					&0.80315\\
learning-to-rank (NN)			&0.80324\\
LambdaMART						&\bf{0.80330}\\
[0.2cm]
aggregate average				&0.80378\\
aggregate RankNet				&\bf{0.80476}\\
\bottomrule
\end{tabular}
\label{tb:results}
\end{table}

After verifying that personalization from logs is possible, we proceeded 
to learning-to-rank and regression models. We trained 1-hidden layer neural
net implementations of each model using tanh activation units and varying the 
number of hidden units in the [10, 200] range. Regression models were optimized 
with squared-loss objective function. Before learning, all features were standardized 
to have mean 0 and standard deviation of 1. For each model we used mini-batch 
learning with batch size of 100 queries (1000 documents), processing each query 
in parallel. Parallel processing allowed us to fully train these models on all 
of the available training data in several hours using the same 
Thinkpad W530 machine.

Results for best neural net (NN) regression and learning-to-rank models
are shown in Table \ref{tb:results}. From the table we see that 
both models significantly improve NDCG@10 with relative gains of up to
0.0118 over the baseline ranking. We also see that regression models
perform similarly to learning-to-rank ones with learning-to-rank only
providing marginal gains. For both types of models we found that neural nets 
with 50 - 100 hidden units performed the best. Moreover, 
for learning-to-rank we found that RankNet performed slightly better 
than other ranking models but the difference was not significant 
(less than 0.0001). 

Best result for LambdaMART is also shown in Table \ref{tb:results}. 
We used publicly available RankLIB library \cite{RankLIB} to run 
LambdaMART. Training LambdaMART took a very long time 
(on the order of days) and used close to 25GB of RAM. We were thus unable 
to properly validate/tune all the hyper parameters such as the number 
of leaves and learning rate. This possibly explains the marginal 
performance of this model as seen from Table \ref{tb:results}, where it 
is performing comparably to the neural net models.

\subsection{Model Aggregation}

For each experiment that we ran throughout the competition we
saved models that performed best on the validation set.
This gave us $\sim$30 trained models at the end of 
the competition. It's well known that blending improves accuracy 
of individual models, and blended solutions have won many data 
mining competitions including the Netflix challenge \cite{Netflix}. 
Keeping this in mind we spent the last few days of the competition 
finding the best blend of the models that we had trained.

Before applying any blending techniques we standardized the 
scores produced by each model to have mean 0 and standard 
deviation 1. After normalization we began with a simple baseline
that averaged all the available scores. This baseline obtained
an NDCG@10 of 0.80378 and is shown in Table \ref{tb:results} 
(``aggregate average"). While this is an improvement over the best
individual model, the improvement is not significant. This
can be attributed to the fact that many models in our blending 
set were considerably weaker than the best model. Consequently, 
including all of these models in the blend with
equal weight significantly affected the overall accuracy. 
It is thus evident that with many weaker models simple 
averaging is not optimal and more adaptive techniques are necessary.
\begin{figure*}[t]
\centerline{
	\subfigure[]{
		\label{fig:kendall}
		\includegraphics[scale=0.36]{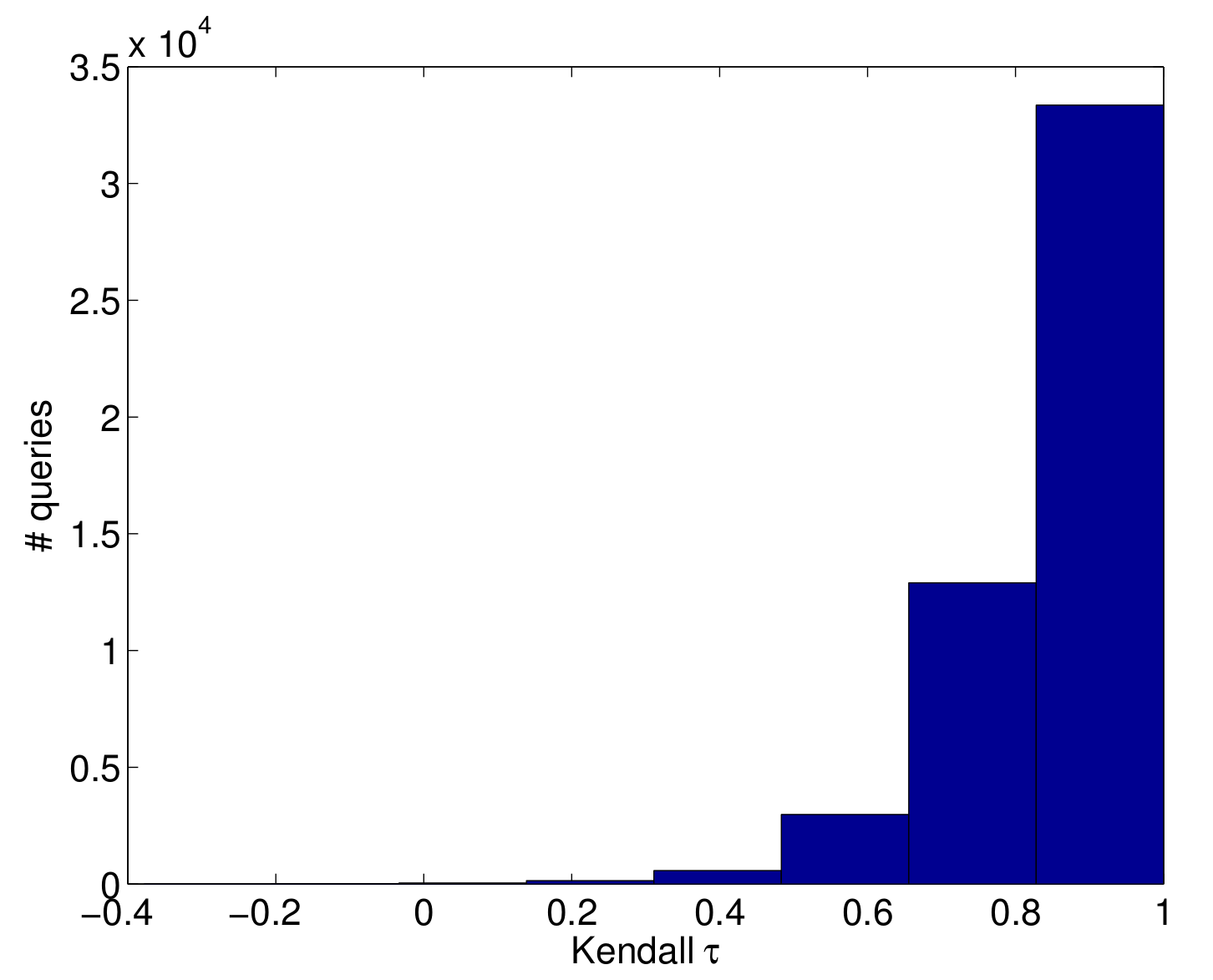}
	}
	\hskip 2cm
	\subfigure[]{
		\label{fig:delta_ndcg}
		\includegraphics[scale=0.36]{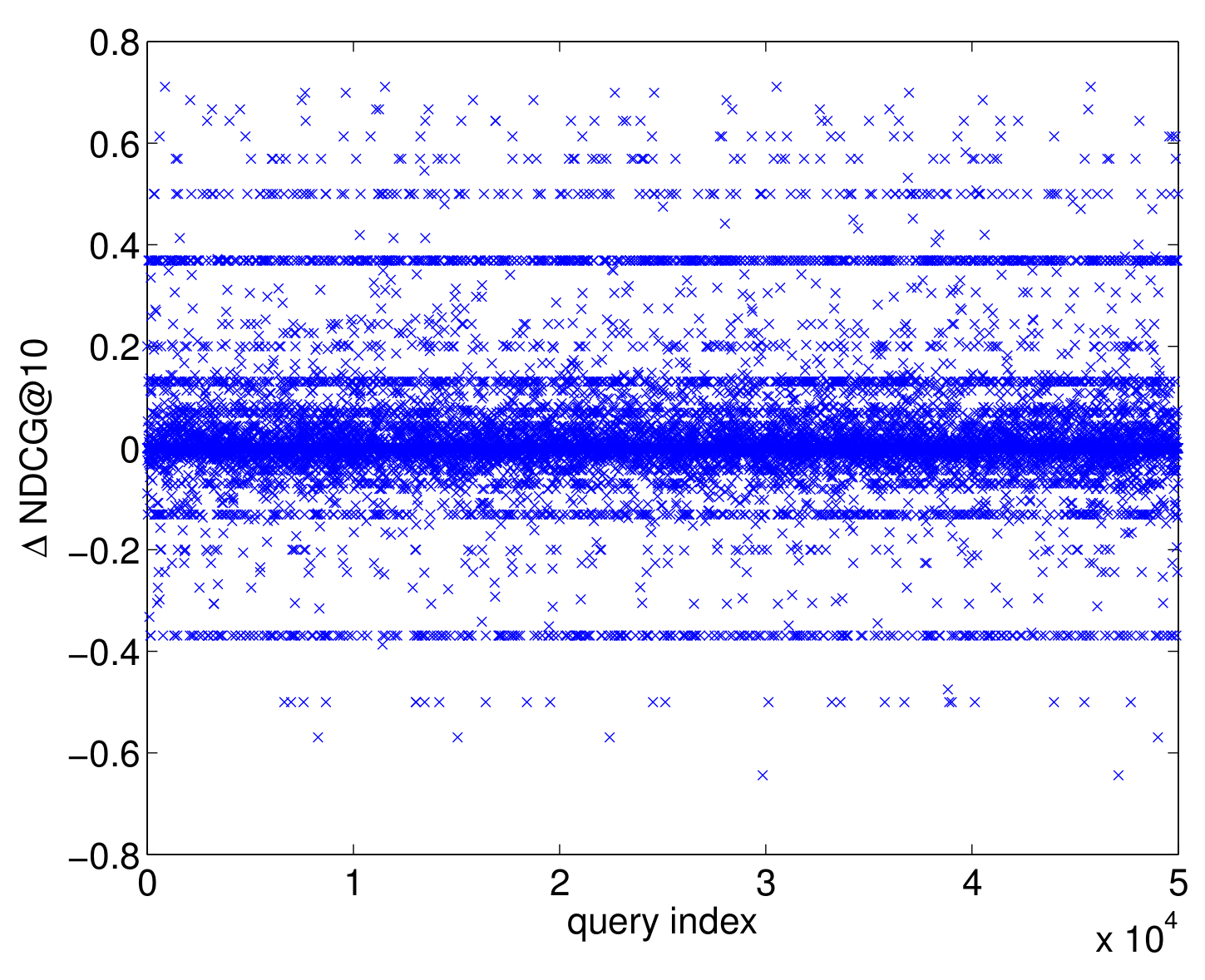}
	}
}
\caption{Figure \ref{fig:delta_ndcg} shows NDCG@10 difference 
between our best personalized model and static ranking produced by Yandex
for 50K validation queries. Figure \ref{fig:kendall} shows Kendall 
$\tau$ distance histogram for the same 50K queries. Kendall $\tau$
is computed between personalized and non-personalized ranking for each
query.}
\label{fig:rerank_plots}
\end{figure*}

One possible solution is to use model-specific 
weights during aggregation. Weights are typically chosen to 
be a function of model's accuracy and several such functions have
have been suggested in literature \cite{WeightedBorda}. 
However, instead of tuning these weights by hand  a 
more principled and potentially more accurate approach is 
to apply one of the learning-to-rank methods to automatically 
learn the weights.

We experimented with this approach and began by partitioning our 
validation\footnote{Note that training set should not be used for 
aggregation since individual models could have overfitted on it.} 
set into two subsets. One subset was then used to train a
linear RankNet on score outputs of all models in the aggregating set, 
and the other subset was used for validation. The result for this 
model is shown at the bottom of Table \ref{tb:results} (``aggregate RankNet"). 
It produced an NDCG@10 of 0.80476 and was our best submission in this 
competition placing 4'th on the private leaderboard.

\subsection{Analysis of Results}

To analyze the effect of personalization we computed Kendall 
$\tau$ correlations between rankings produced by
our best model and the non-personalized baseline rankings from Yandex.
The plot for randomly chosen 50K validation queries is 
shown in Figure \ref{fig:kendall}. From this figure we see that 
for most queries $\tau$ is above 0.7 indicating that
our model is fairly conservative and tends to only re-rank a 
few documents in the list. However, we also see that a number of
queries are very aggressively re-ranked with $\tau$ below 0.5.

While aggressive personalization can significantly improve
user search experience, it can also lead to dangerous outlier queries
where top-N documents are ranked completely out of order. This is
further illustrated in Figure \ref{fig:kendall} which shows the
difference in NDCG@10 between our model and Yandex's base
ranking for the same 50K queries. From this figure we see that 
while personalized model improves NDCG for many queries, some queries 
are also significantly hurt with NDCG drops of over 0.4. This further 
demonstrates the danger of applying personalization 
to all queries and emphasizes the need for adaptive strategies 
that selectively choose which queries should be re-ranked. 
Moreover, risk minimization (largest NDCG loss across all queries)
might be a more appropriate objective for this task since it 
can produce models with more stable worst-case performance.
This, however, is beyond the scope of this paper 
and we leave it for future research.

\section{Conclusion and Future Work}

In this paper we presented our solution to the Yandex Personalized 
Web Search Challenge. In our approach search logs were first partitioned 
into user and query dependent neighborhoods (contexts). Query-document
features were then extracted from each context summarizing document 
preference within the context. Models trained on these features achieved 
significant improvements in accuracy over non-personalized ranker.

In the future work we plan to explore contexts based on {\it similar}
queries/users. Such contexts have been successfully applied in 
neighborhood-based collaborative filtering models and can potentially
be very useful in this domain as well. Both user an query
similarities can be readily inferred from the search logs using
statistics like issued query overlap for users and document/domain overlap
for queries. These contexts can be particularly useful for personalization
of long-tail queries that occur very infrequently in the data and do
not have enough preference data.

\section{Acknowledgments}

We would like to thank Yandex for organizing this competition 
and for releasing such an interesting large-scale dataset.
We would also like to thank Kaggle for hosting and managing 
the competition.

\bibliographystyle{abbrv}
\bibliography{yandex_kaggle_model}

\appendix

\section{Context Features}\label{app:context_features}

Given user-query-document triplet $(u, q_u, d_{q_u})$ and context 
$\C$ we extract a total of 20 context-dependent features 
$g_1$ - $g_{20}$ (all missing features are set to 0):
\begin{itemize}
\item Total relevance for all clicks on $d_{q_u}$ in $\C$:
\begin{equation*}
g_1 = \sum_{q \in \C} \sum_{d_{q} \in \D_q} I[d_{q} = d_{q_u}] l_{q}
\end{equation*}
where $I[x]$ is an indicator function evaluating to 1 if $x$ is true and
0 otherwise

\item Average relevance for all clicks on $d_{q_u}$ in $\C$:
\begin{equation*}
g_2 = \frac{1}{\sum_{q \in \C} \sum_{d_{q} \in \D_q} I[d_{q} = d_{q_u}]} 
\sum_{q \in \C} \sum_{d_{q} \in \D_q} I[d_{q} = d_{q_u}] l_{q}
\end{equation*}

\item Max/min relevance across all clicks on $d_{q_u}$ in $\C$:
\begin{equation*}
g_3 = \arg \max \{l_{q} | q \in \C, d_{q} \in \D_q, d_{q} = d_{q_u}\}
\end{equation*}
\begin{equation*}
g_4 = \arg \min \{l_{q} | q \in \C, d_{q} \in \D_q, d_{q} = d_{q_u}\}
\end{equation*}

\item Average similarity between $q_u$ and all queries in $\C$ where
$d_{q_u}$ was clicked:
\begin{equation*}
g_5 = \frac{1}{\sum_{q \in \C} \mbox{clicked}(d_{q_u}, \D_q)} 
\sum_{q \in \C} \mbox{clicked}(d, \D_q) \mbox{sim}(q, q_u)
\end{equation*}
where $\mbox{clicked}(d_{q_u}, \D_q) = 1$ if $d$ was clicked in $\D_q$ and 0 otherwise.
$\mbox{sim}(q, q_u)$ is similarity between $q$ and $q_u$, in this work we use
intersection over union metric applied to query terms.

\item Max similarity between $q_u$ and all queries in $\C$ where
$d_{q_u}$ was clicked:
\begin{equation*}
g_6 = \arg \max \{\mbox{sim}(q, q_u) | q \in \C, \mbox{clicked}(d_{q_u}, \D_q) = 1\}
\end{equation*}

\item Average similarity between $q_u$ and all queries in $\C$ where
$d_{q_u}$ was skipped (i.e. $d_{q_u}$ was not clicked but there was at least on click
below $d_{q_u}$):
\begin{equation*}
g_7 = \frac{1}{\sum_{q \in \C} \mbox{skipped}(d_{q_u}, \D_q)} 
\sum_{q \in \C} \mbox{skipped}(d_{q_u}, \D_q) \mbox{sim}(q, q_u)
\end{equation*}
where $\mbox{skipped}(d_{q_u}, \D_q) = 1$ if $d_{q_u}$ was skipped in $\D_q$ and 0 
otherwise.

\item Max similarity between $q_u$ and all queries in $\C$ where
$d_{q_u}$ was skipped:
\begin{equation*}
g_8 = \arg \max \{\mbox{sim}(q, q_u) | q \in \C, \mbox{skipped}(d_{q_u}, \D_q) = 1\}
\end{equation*}

\item Average similarity between $q_u$ and all queries in $\C$ where
$d_{q_u}$ was missed (i.e. all clicks were above $d$):
\begin{equation*}
g_9 = \frac{1}{\sum_{q \in \C} \mbox{missed}(d_{q_u}, \D_q)} 
\sum_{q \in \C} \mbox{missed}(d_{q_u}, \D_q) \mbox{sim}(q, q_u)
\end{equation*}
where $\mbox{missed}(d_{q_u}, \D_q) = 1$ if $d_{q_u}$ was missed in $\D_q$
and 0 otherwise.

\item Max similarity between $q_u$ and all queries in $\C$ where
$d_{q_u}$ was missed:
\begin{equation*}
g_{10} = \arg \max \{\mbox{sim}(q, q_u) | q \in \C, \mbox{missed}(d_{q_u}, \D_q) = 1\}
\end{equation*}

\item Number of times $d_{q_u}$ was shown, clicked, skipped and missed in $\C$:
\begin{equation*}
g_{11} = \sum_{q \in \C} I[d_{q_u} \in \D_q]
\end{equation*}
\begin{equation*}
g_{12} = \sum_{q \in \C} \mbox{clicked}(d_{q_u}, \D_q)
\end{equation*}
\begin{equation*}
g_{13} = \sum_{q \in \C} \mbox{skipped}(d_{q_u}, \D_q) 
\end{equation*}
\begin{equation*}
g_{14} = \sum_{q \in \C} \mbox{missed}(d_{q_u}, \D_q)
\end{equation*}

\item Number of times $d_{q_u}$ was shown in $\C$ discounted by rank:
\begin{equation*}
g_{15} = \sum_{q \in \C} \frac{1}{\mbox{r\_shown}(d_{q_u}, \D_q)}
\end{equation*}
where $\mbox{r\_shown}(d_{q_u}, \D_q)$ is rank of $d_{q_u}$ in $\D_q$ if
it was shown and 0 otherwise. When $\mbox{r\_shown}(d_{q_u}, \D_q) = 0$
the ratio is set to 0.

\item Number of times $d_{q_u}$ was clicked in $\C$ discounted by rank:
\begin{equation*}
g_{16} = \sum_{q \in \C} \frac{1}{\mbox{r\_clicked}(d_{q_u}, \D_q)}
\end{equation*}
where $\mbox{r\_clicked}(d_{q_u}, \D_q)$ is rank of $d_{q_u}$ in $\D_q$ if
it was clicked and 0 otherwise. When $\mbox{r\_shown}(d_{q_u}, \D_q) = 0$
the ratio is set to 0.

\item Max/min rank of $d_{q_u}$ when it was clicked in $C$
\begin{equation*}
g_{17} = \arg \max \{\mbox{r\_clicked}(d_{q_u}, \D_q) | q \in \C\}
\end{equation*}
\begin{equation*}
g_{18} = \arg \min \{\mbox{r\_clicked}(d_{q_u}, \D_q) | q \in \C\}
\end{equation*}

\item Number of times $d_{q_u}$ was skipped in $\C$ discounted by rank:
\begin{equation*}
g_{19} = \sum_{q \in \C} \frac{1}{\mbox{r\_skipped}(d_{q_u}, \D_q)}
\end{equation*}
where $\mbox{r\_clicked}(d_{q_u}, \D_q)$ is rank of $d_{q_u}$ in $\D_q$ if
it was skipped and 0 otherwise. When $\mbox{r\_skipped}(d_{q_u}, \D_q) = 0$
the ratio is set to 0.

\item Number of times $d_{q_u}$ was missed in $\C$ discounted by rank:
\begin{equation*}
g_{20} = \sum_{q \in \C} \frac{1}{\mbox{r\_missed}(d_{q_u}, \D_q)}
\end{equation*}
where $\mbox{r\_clicked}(d_{q_u}, \D_q)$ is rank of $d_{q_u}$ in $\D_q$ if
it was missed and 0 otherwise. When $\mbox{r\_missed}(d_{q_u}, \D_q) = 0$
the ratio is set to 0.

\end{itemize}

\end{document}